\pdfoutput=1
\documentclass[11pt]{article}
\usepackage{booktabs} 
\usepackage[table,xcdraw]{xcolor} 
\usepackage{multirow} 
\usepackage{array} 
\usepackage[ruled,vlined]{algorithm2e}
\usepackage{adjustbox}
\usepackage{amsmath}
\usepackage[preprint]{coling}

\usepackage{times}
\usepackage{latexsym}

\usepackage[T1]{fontenc}

\usepackage[utf8]{inputenc}

\usepackage{microtype}

\usepackage{inconsolata}

\usepackage{graphicx}

\title{AutoProteinEngine: 
A Large Language Model Driven Agent Framework for Multimodal AutoML in Protein Engineering
}

\author{
 \textbf{Yungeng Liu\textsuperscript{1,2,$\dagger$}},
 \textbf{Zan Chen\textsuperscript{1,$\dagger$}},
 \textbf{Yu Guang Wang\textsuperscript{1,3}},
 \textbf{Yiqing Shen\textsuperscript{4,*}},
\\
\\
 \textsuperscript{1}\textit{Toursun Synbio}, Shanghai, China 
 \textsuperscript{2}\textit{City University of Hong Kong}, Hong Kong SAR\\
 \textsuperscript{3}\textit{Shanghai Jiao Tong University}, Shanghai, China
 \textsuperscript{4}\textit{Johns Hopkins University}, Baltimore, USA
 \\
\\
 \small{
    \textbf{\textsuperscript{$\dagger$}Equal Contribution.}
   \textbf{\textsuperscript{*}Corresponding Author:} Yiqing Shen (\href{yiqing.shen1@gmail.com}{yiqingshen1@gmail.com}).
 }
}

\begin{document}
\maketitle
\begin{abstract}
Protein engineering is important for biomedical applications, but conventional approaches are often inefficient and resource-intensive. 
While deep learning (DL) models have shown promise, their training or implementation into protein engineering remains challenging for biologists without specialized computational expertise. 
To address this gap, we propose AutoProteinEngine (AutoPE), an agent framework that leverages large language models (LLMs) for multimodal automated machine learning (AutoML) for protein engineering.
AutoPE innovatively allows biologists without DL backgrounds to interact with DL models using natural language, lowering the entry barrier for protein engineering tasks. 
Our AutoPE uniquely integrates LLMs with AutoML to handle model selection for both protein sequence and graph modalities, automatic hyperparameter optimization, and automated data retrieval from protein databases.
We evaluated AutoPE through two real-world protein engineering tasks, demonstrating substantial performance improvements compared to traditional zero-shot and manual fine-tuning approaches. 
By bridging the gap between DL and biologists' domain expertise, AutoPE empowers researchers to leverage DL without extensive programming knowledge. 
Our code is available at \url{https://github.com/tsynbio/AutoPE}.

\end{abstract}

\section{Introduction}
Protein engineering, focused on designing and optimizing proteins with enhanced and tailored functions, plays a crucial role in a wide range of  biomedical applications including drug discovery, enzyme optimization, and biomaterial design \cite{brannigan2002protein, carter2011introduction}.
Traditional approaches to protein engineering, such as directed evolution and rational design, are often constrained by inefficiency, low success rates, and high resource demands \cite{goldsmith2012directed}. 
Deep learning (DL) models, such as the ESM series \cite{lin2023evolutionary, verkuil2022language, rives2021biological} and AlphaFold series \cite{jumper2021highly, evans2021protein} models, have enhanced efficiency and accuracy of protein structure prediction, understanding protein-protein interactions, and other tasks within protein engineering.
However, training or fine-tuning those deep learning models for specific protein engineering tasks poses significant challenges for biologists lacking specialized coding and machine learning expertise \cite{yang2019machine}.
Specifically, the intricate architectures of deep learning models require a deep understanding of DL principles for effective interpretation and modification. 
Optimizing model performance further necessitates adjusting hyperparameters, a process that relies heavily on machine learning experience and intuition.
Moreover, preparing protein data for input into these models often involves specialized pre-processing techniques.
Finally, the complexity is amplified
by the multimodal nature of protein data, which can be represented in both sequence and protein graph formats, adding an additional layer of difficulty of model training and optimization.

Although automated machine learning (AutoML)  \cite{waring2020automated} has been introduced to reduce the manual effort involved in training DL models \cite{xiao2022passer2, chen2021ilearnplus}, existing AutoML frameworks still demand considerable expertise in DL and programming. This limits their accessibility to biologists who lack computational backgrounds \cite{luo_autom3l_2024}. 
Moreover, these frameworks are typically designed for general tasks and therefore lack domain-specific knowledge of protein engineering, limiting their capability to handle protein sequences and protein graphs.
To address these challenges, we propose an agent framework that leverages large language models (LLMs) for multimodal AutoML specifically tailored to protein engineering.
LLMs offer the advantage of interacting with the model in a conversational manner, which can reduce the learning curve for users \cite{zhang2024exploring,shen2024toursynbio}. 
Our approach aims to bridge the gap between DL models and biologists' domain expertise, enabling more efficient and accessible protein engineering workflows while incorporating the necessary domain-specific knowledge for handling protein data across various modalities.

The major contributions of this work are three-fold.
Firstly, we propose an innovative LLM-based agent framework for multimodal AutoML specifically for protein engineering tasks, namely \textbf{AutoProteinEngine} (AutoPE).
To the best of our knowledge, this is the first attempt at a multimodal AutoML framework for protein engineering, that can tackle both the protein sequence and protein graph.
Notably, AutoPE allows users to perform AutoML tasks through conversational interactions with the framework.
Secondly, to further boost the performance, we propose an automated hyper-parameter optimization module that conducts hyper-parameter search via LLM.
Finally, we propose an automated data retrieval method that can facilitate seamless data retrieval from protein databases such as PDB, and UniProt, using natural language descriptions.

\begin{figure*}[t!]
  \centering
  \includegraphics[width=0.95\linewidth]{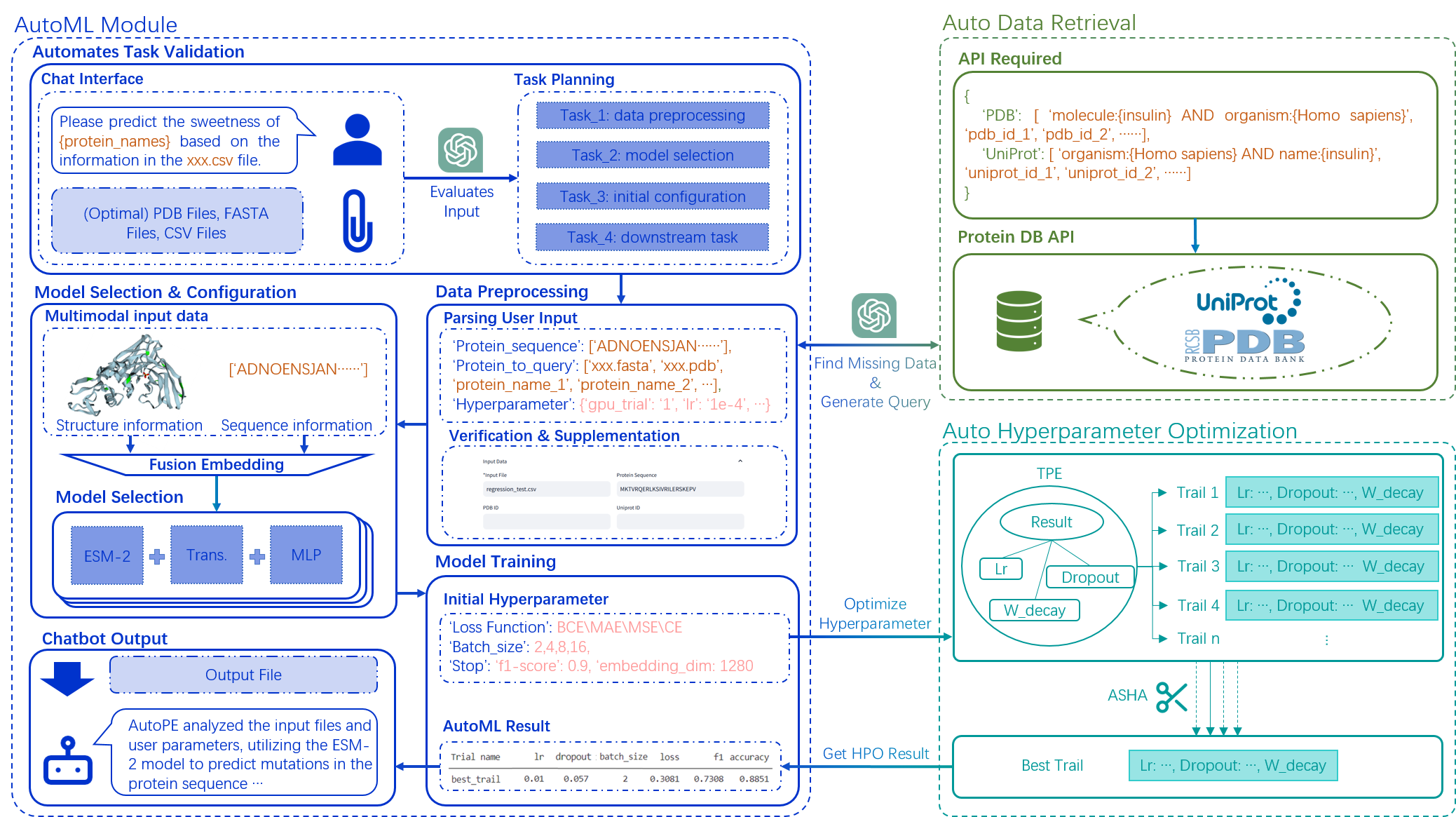} \hfill
  \caption {
  The overview of AutoProteinEngine (AutoPE) framework. 
  It illustrates the end-to-end workflow of AutoPE, integrating LLM-driven AutoML for protein engineering tasks.
  The framework consists of three main components: (1) A user-friendly chat interface for the natural language task specification and data input; (2) An AutoML module that handles task validation, data preprocessing, model selection, and an auto HPO module that automizes the HPO searching; and (3) An auto data retrieval module for acquiring protein-related data from databases like UniProt and PDB. 
  It supports multimodal protein data (protein sequences and structures) and provides interactive feedback throughout the workflow. 
  }
  \label{fig:framework}
\end{figure*}

\section{Methods}
\subsection{LLM-driven AutoML}
At the core of AutoPE is its AutoML module, which features a core AutoML module that automates task validation, data preprocessing, model selection and configuration, and model training for protein engineering tasks \cite{erickson2020autogluon}. 
This module is driven by an LLM, enabling a user-friendly interface where users (\textit{e}.\textit{g}., biologists) without extensive computational expertise can specify their tasks using natural language.

Users describe their protein engineering task in natural language, such as ``\textit{I need to train a model to predict the mutation of given protein sequence}''.
The LLM then evaluates this input to determine if it aligns with AutoPE's capabilities. 
This task validation stage is achieved through a prompt that includes context about protein engineering tasks and AutoPE's functionalities. 
Specifically, the LLM is instructed to determine if the task falls within valid categories depending on the available model zoo for the model selection stage, such as protein stability prediction, protein-protein interaction prediction, enzyme activity prediction, or protein mutation. 
If the task is not immediately clear or outside AutoPE's scope, it engages in a dialogue to clarify or refine the request.

Once a task is validated, the LLM analyzes the input to formulate a plan before action. 
This plan encompasses data preprocessing strategies, model selection, and configuration stage that selects models from predefined model zoos (\textit{e}.\textit{g}., ESM series, AlphaFold).
The LLM generates this using retrieve augmented generation (RAG) on related literature, incorporating domain-specific protein engineering knowledge to ensure all aspects of the task are addressed.

In the data preprocessing stage, if the input data is incomplete, AutoPE's data retrieval module comes into play. 
It supplements the data by accessing online sources including PDB (Protein Data Bank) and UniProt databases. 
This process is also guided by the LLM, which formulates appropriate database queries based on the task requirements.

With a complete dataset and a formulated plan, AutoPE proceeds to the model selection and configuration stage.
The AutoPE executes the plan by selecting and configuring appropriate models from the predefined model zoos. 
For tasks involving multimodal data, which include both sequence and graph representations of proteins, AutoPE implements a late fusion scheme.
Specifically, it combines embeddings from different modalities, allowing AutoPE to leverage complementary information from both sequence and structural data.

In the training stage, the LLM is prompted to refine a pre-defined general training framework by incorporating model-specific optimizations. 
This includes selecting appropriate loss functions, determining optimal batch sizes and learning rates based on the selected model and dataset size, and implementing early stopping and model check-pointing to prevent overfitting. 
The LLM also applies task-specific data augmentation techniques, such as random mutations for sequence data or graph perturbations for structural data. 
For Transformer-based models like ESM, it fine-tunes attention mechanisms or the prediction head depending on the task types.
The overall framework of AutoPE is depicted in Fig. \ref{fig:framework}.

\subsection{Auto Hyperparameter Optimization}
To further enhance the performance and usability of AutoPE, we introduce an auto hyperparameter optimization (HPO) module that enables HPO searching via natural language guidance from the user and provides better interpretations of results. 
The core of the Auto HPO module comprises two stages, namely the Tree-structured Parzen Estimator (TPE) and the Asynchronous Successive Halving Algorithm (ASHA) \cite{watanabe2023tree, li2018massively}. 
The TPE algorithm optimizes hyperparameter configurations by modeling the probability of a configuration yielding good performance, defined by
%
$x^* = \arg \max_{x} \frac{l(x)}{g(x)}$,
where $x^*$ represents the optimal hyperparameter configuration, $l(x)$ and $g(x)$ are the likelihood functions for high and low-performing configurations respectively.
The TPE approach allows Autope to efficiently explore the hyperparameter space, focusing on regions that are more likely to yield improved performance.
Complementing TPE, the ASHA scheduler optimizes resource allocation through a multi-fidelity approach, described by:
\begin{equation}
\begin{aligned}
r_i &= r_{\min} \cdot \eta^i \\
n_i &= \left\lfloor \frac{n}{\eta^i} \right\rfloor \\
T &= \sum_{i=0}^{\lfloor \log_\eta(n) \rfloor} n_i \cdot r_i,
\end{aligned}
\end{equation}
where, $r_i$ denotes the resource allocation at the $i$-th iteration, $n_i$ represents the number of configurations evaluated, and $T$ is the total computational budget. 
ASHA allows for early termination of underperforming experiments, dynamically reallocating resources to more promising configurations.

\begin{figure*}[t]
\centering
\centerline{\includegraphics[width=\linewidth]{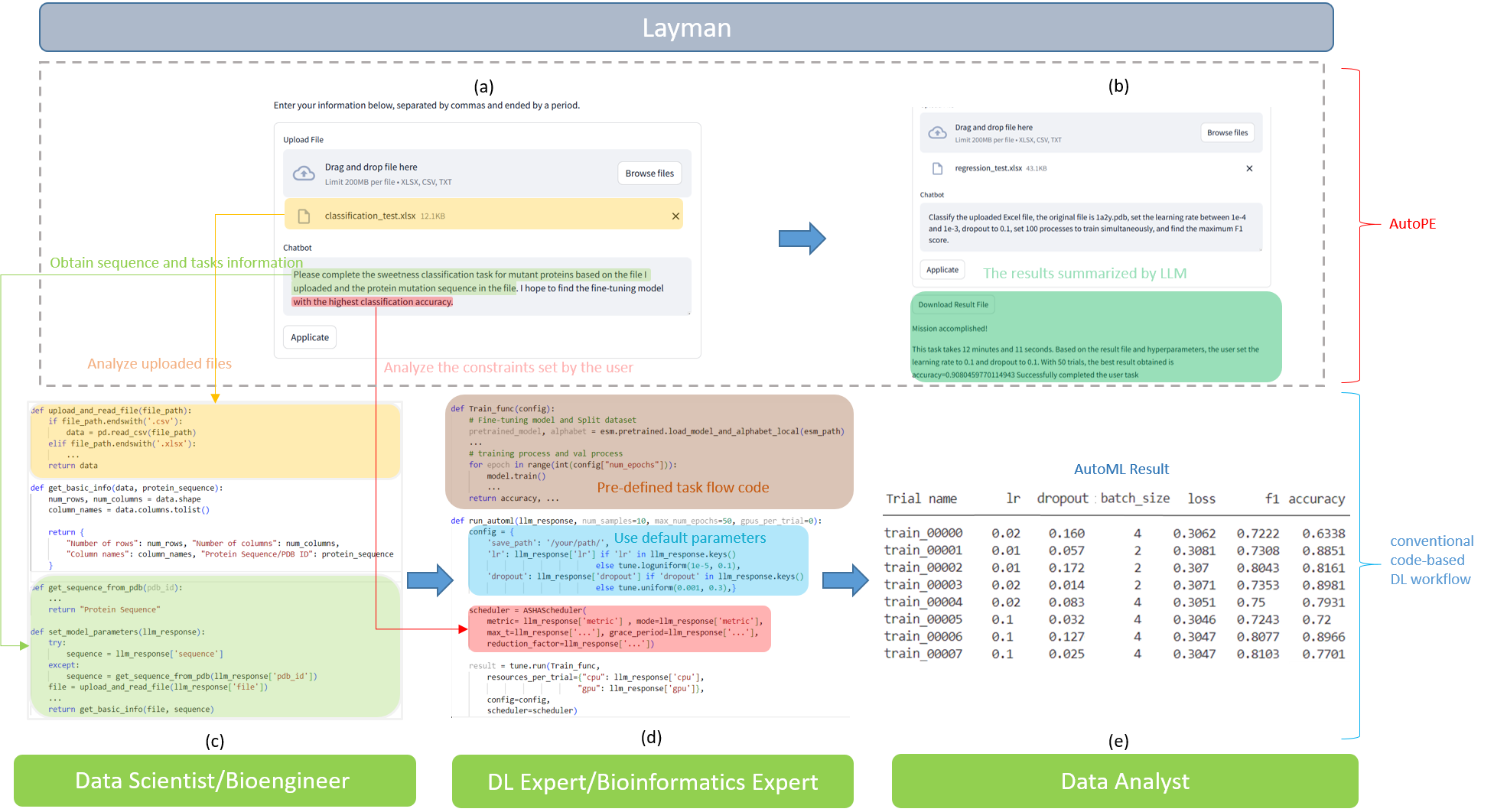}}
  \caption {
Case study between the AutoPE in a conversational interface with conventional code-based DL workflow for brazzein protein sweetness classification task. 
This figure demonstrates the end-to-end process and improved usability of AutoPE: (a) A biologist without DL background uploads protein mutation data and specifies the sweetness classification task using natural language; (b) LLM-driven analysis of inputs and automated result interpretation, showing AutoPE's ability to handle domain-specific requests.
In the conventional code-based DL workflow multiple experts are involved: (c) Pre-defined modules process the Brazzein protein sequences require both data scientist and biologist; (d) AutoML pipeline, including HPO, optimizes the classification model requires programming experts; (e) Results visualization requires data analyst. 
This case study highlights AutoPE's effectiveness in enabling non-expert users to perform complex protein engineering tasks, achieving superior performance while requiring minimal technical expertise.
}
  \label{fig:casestudy}
\end{figure*}

AutoPE utilizes \texttt{Ray.Tune} \cite{liaw2018tune} to manage the HPO process, where the LLM summarizes and verifies user inputs before initiating the optimization.
Before HPO, AutoPE interacts with the user to confirm hyperparameter settings or suggest additional configurations, which allows researchers to leverage their domain knowledge while benefiting from the LLM's ability to navigate complex hyperparameter spaces.
During the HPO process, AutoPE provides feedback throughout the optimization process, which communicates progress and results, converting numerical metrics (such as MSE or F1-score) into user-friendly natural language summaries, as shown in Fig. \ref{fig:casestudy}.
It enhances the interpretability of the optimization process, allowing users to gain insights into the performance trends of different hyperparameter configurations.
Resource management during hyperparameter optimization is similarly facilitated via LLM interaction. Users can specify computational preferences, such as the number of GPUs to allocate to each trial, through natural language commands.

\subsection{Auto Data Retrieval}
AutoPE's auto data retrieval module streamlines the acquisition of essential protein-related data for pretraining and fine-tuning models in protein engineering tasks.
It provides a user-friendly design for specifying data requirements through natural language queries.
Users can request data in various formats, including protein sequences, PDB structures, UniProt IDs, or general protein descriptions, where LLM interprets these requests.
For example, if a user inputs ``\textit{I need the sequence and structure data for human insulin}'', the LLM processes this natural language query and translates it into specific data retrieval tasks. 
Specifically, it identifies key elements such as the protein name (insulin), the organism (human), and the required data types (sequence and structure).
Upon parsing the user's request, the auto data retrieval module leverages the LLM to construct appropriate database queries. 
For UniProt \cite{uniprot2019uniprot}, the LLM generates a query like ``organism:\{Homo sapiens\} AND name:\{insulin\}''. 
Similarly, for PDB \cite{burley2017protein}, it constructs a query such as ``molecule:\{insulin\} AND organism:\{Homo sapiens\}''. 
The LLM's ability to generate these structured queries from natural language input enables efficient and accurate data retrieval across multiple databases.
In cases where data is unavailable or incomplete, AutoPE engages in an interactive dialogue guided by the LLM. 
If the initial search yields no results, the LLM further prompts the user with alternative options, such as searching for closely related insulin structures from other mammals or focusing on sequence data only.

This module also enhances data retrieval flexibility by allowing users to manually input or verify the retrieved data through an editable table interface, which is particularly useful for incorporating proprietary or unpublished private data that may not be available in public databases. 
The LLM assists in this process by providing guidance on data formatting and validating user inputs to ensure consistency with the required data structure for downstream analyses.
After compiling the necessary data, AutoPE presents users with a detailed summary of the collected information and the overall task scope. 
This summary, generated by the LLM, includes a list of retrieved protein sequences and their sources, PDB IDs of relevant structures and their resolution, key UniProt annotations for the proteins of interest, and any gaps or potential issues in the collected data.

\section{Experiments}

\subsection{Datasets}
To evaluate our AutoPE framework, we selected two distinct proteins for classification and regression tasks, respectively.
For the classification task, we focused on Brazzein, a high-intensity sweetener protein originally isolated from the West African plant Pentadiplandra brazzeana \cite{ming1994brazzein, assadi2005brazzein}. 
The dataset consists of 435 mutation entries, comprising single and multi-point mutations of Brazzein protein, along with their corresponding relative sweetness measurements. 
We categorized the mutations as ``sweet'' or ``non-sweet'' based on a threshold relative sweetness of 100 (equivalent to sucrose).
The regression task utilized data from the STM1221 wild-type protein, an enzyme that specifically removes acetyl groups from target proteins. 
Our dataset consisted of 234 enzyme activity scores for various mutation scenarios, as determined through wet lab experiments. 
This continuous data enables the prediction of enzyme activity levels based on specific mutations.
We randomly partitioned each dataset into training (80\%) and testing (20\%) sets with a five-fold validation.

\subsection{Experiment Design and Implementation}%
We designed our experiments to compare the performance of AutoPE with two distinct approaches, namely zero-shot inference and manual fine-tuning. 

\paragraph{Zero-Shot Inference}
The zero-shot inference leverages pre-trained protein language models \textit{i}.\textit{e}., ESM to extract features without task-specific fine-tuning. 
This approach provides a baseline for assessing the generalization capabilities of AutoPE. 
After extracting feature representations, we leverage traditional machine learning algorithms such as logistic regression and k-nearest neighbors (KNN) for functional score prediction. 
Grid search is employed to identify the optimal algorithm and hyperparameters, with the final performance evaluated on the test set.

\paragraph{Manual Fine-Tuning}
To enhance model performance on specific tasks, we manually fine-tune the pre-trained protein language models. 
Our customization includes adding a self-attention layer to capture medium and long-range dependencies in protein sequences, improving the model's ability to identify complex mutations. 
This setting aims to enhance feature capture and non-linear transformation capabilities, potentially leading to improved task-specific performance.
We also manually searched HPO by a well-experienced ML engineer, who has three years of experience in both DL and protein engineering. 



\paragraph{Evaluation Metrics}
For classification, we utilize three metrics,
namely (1) the F1-score that balances precision and recall and is important for imbalanced datasets; (2) ROC-AUC (Receiver Operating Characteristic - Area Under Curve), which evaluates the model's overall discriminative ability across various thresholds, with higher values indicating superior performance; (3) Spearman Rank Correlation Coefficient (SRCC) that assesses rank preservation. 
For regression, we employ another three metrics, namely (1) Mean Squared Error (MSE) which quantifies the average squared deviation between predicted and true values; (2) Mean Absolute Error (MAE), which complements MSE by calculating the average absolute deviation with reduced sensitivity to outliers;
(3) R\textsuperscript{2} score that evaluates the model's explanatory power by measuring the proportion of variance in the dependent variable accounted for by the model. 
A higher R\textsuperscript{2} score, approaching 1, indicates better alignment between predictions and ground truth. 

\paragraph{Implementation Details}
For the zero-shot method, we explored multiple machine learning algorithms, including Support Vector Machines (SVM), Random Forest (RF), and Logistic Regression. For SVM, we tested linear, radial basis function (RBF), and polynomial kernels. The Random Forest classifier was evaluated with various hyperparameters: number of estimators (50, 100, 200), maximum depth (None, 10, 20, 30), and minimum samples split (2, 5, 10). 
For manual fine-tuning, we employed a consistent set of hyperparameters for initialization: dropout rate of 0.30, learning rate of 1e-3, batch size of 8, and 50 training epochs. Weight decay was set to 1e-5 to prevent overfitting. We also implemented a learning rate scheduler with a step size of 10 and a $\gamma$ of 0.1 to gradually reduce the learning rate during training.
For the LLM used in AutoPE, we utilize the TourSynbio-7B \cite{shen2024toursynbio} due to its outstanding performance on protein understanding.
All implementations are conducted on 8 NVIDIA 4090 GPU cards.

\subsection{Results}

\begin{table}[!htbp]
\centering
\caption{Performance comparison to zero-shot inference and manual fine-tuning on Brazzein protein sweetness classification task.}
\resizebox{\columnwidth}{!}{%
\begin{tabular}{lccc}
    \toprule
    \rowcolor[HTML]{EFEFEF} 
    \textbf{Methods} & \textbf{F1-score ↑} & \textbf{SRCC ↑} & \textbf{Accuracy ↑} \\
    \midrule
    \rowcolor[HTML]{E6F7FF} 
    Zero-Shot & 0.4764 \footnotesize{± 0.11} & 0.3769 \footnotesize{± 0.05} & 0.6917 \footnotesize{± 0.04} \\

    \rowcolor[HTML]{E6F7FF} 
    Manual Fine-Tuning & 0.5709 \footnotesize{± 0.05} & 0.3098 \footnotesize{± 0.06} & \textbf{0.9137} \textbf{\footnotesize{± 0.01}} \\
    
    \rowcolor[HTML]{FFF5E6} 
    \textbf{AutoPE (w/o HPO)} & 0.6396 \footnotesize{± 0.06} & 0.4405 \footnotesize{± 0.04} & 0.7988 \footnotesize{± 0.05} \\
    \rowcolor[HTML]{FFF5E6} 
    \textbf{AutoPE (w/ HPO)} & \textbf{0.7306} \textbf{\footnotesize{± 0.04}} & \textbf{0.4621} \textbf{\footnotesize{± 0.03}} & 0.8908 \footnotesize{± 0.01} \\
    \bottomrule
\end{tabular}%
}
\label{tab:class_result}
\end{table}

\begin{table}[!b]
\centering
\caption{Performance comparison to zero-shot inference and manual fine-tuning on STM1221 enzyme activity regression task.}
\resizebox{\columnwidth}{!}{%
\noindent
\begin{tabular}{lccc}
    \toprule
    \rowcolor[HTML]{EFEFEF} 
    \textbf{Methods} & \textbf{RMSE ↓} & \textbf{MAE ↓} & \textbf{R2\_score ↑} \\
    \midrule
    \rowcolor[HTML]{E6F7FF}
    Zero-Shot & 0.4862 \footnotesize{± 0.14} & 0.2766 \footnotesize{± 0.15} & 0.5663 \footnotesize{± 0.04} \\
\rowcolor[HTML]{E6F7FF}     
Manual Fine-Tuning & 0.3579 \footnotesize{± 0.15} & 0.2236 \footnotesize{± 0.16} & 0.5965 \footnotesize{± 0.07} \\
    \rowcolor[HTML]{FFF5E6}
    \textbf{AutoPE (w/o HPO)} & 0.4029 \footnotesize{± 0.19} & 0.2164 \footnotesize{± 0.14} & 0.6153 \footnotesize{± 0.09} \\

    \rowcolor[HTML]{FFF5E6}
    \textbf{AutoPE (w/ HPO)} & \textbf{0.3488} \textbf{\footnotesize{± 0.19}} & \textbf{0.1999} \textbf{\footnotesize{± 0.13}} & \textbf{0.6805} \textbf{\footnotesize{± 0.09}} \\
    \rowcolor[HTML]{E6F7FF}
    \bottomrule
\end{tabular}
}
\label{tab:regression_result}
\end{table}

\begin{figure}[t!]
    \centering
    {
        \includegraphics[width=\linewidth]{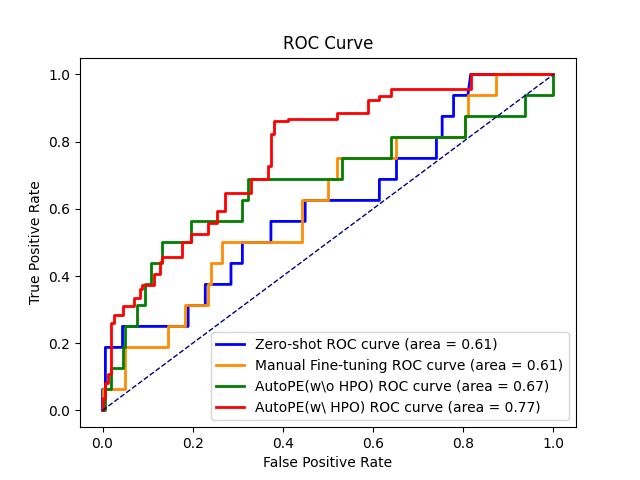}
        \label{fig:roc}
    }
    \caption{ROC curves for zero-shot, manual fine-tuning, and AutoPE on Brazzein protein sweetness classification task.}
    \label{fig:classification_metrics}
\end{figure}

In the classification task (Fig. \ref{fig:classification_metrics}, Tab. \ref{tab:class_result}), AutoPE demonstrated superior performance across all metrics, with its ROC curve closest to the top-left corner. 
The ablation study further underscored AutoPE's efficacy, particularly with the auto HPO module. 
AutoPE with auto HPO module achieved the highest F1 Score (0.7306) and SRCC (0.4621), outperforming both its without auto HPO module variant (F1 score: 0.6396, SRCC: 0.4405) and baselines. 
While manual fine-tuning achieved the highest accuracy, its lower F1 score suggests potential overfitting. 
In contrast, AutoPE with auto HPO module achieves an optimal balance, combining high accuracy (0.8908) with the best F1 score and SRCC, indicating enhanced robustness and generalizability. 
In the regression task (Tab. \ref{tab:regression_result}), AutoPE also demonstrated superior performance across all metrics. 
%
%
AutoPE with auto HPO module achieved the lowest RMSE (0.3488) and MAE (0.1999), surpassing both its non-HPO variant (RMSE: 0.4029, MAE: 0.2164) and baseline methods. 
Notably, AutoPE with auto HPO module attained the highest R2 score (0.6805), indicating superior explanatory power for the variance in the target variable. 
While manual fine-tuning showed competitive performance in RMSE (0.3579), AutoPE with auto HPO module consistently outperformed across all metrics. 
Finally, we perform a case study to show the improved usability of AutoPE in Fig. \ref{fig:casestudy}.

\section{Conclusion}
The AutoProteinEngine (AutoPE) effectively bridges the gap between DL models and biologists' domain expertise. 
By leveraging LLM for multimodal AutoML, AutoPE demonstrates substantial advantages in accessibility, efficiency, and performance. 
It simplifies AutoML task customization and data processing, enabling biologists without extensive computational backgrounds to leverage advanced DL models in protein engineering tasks. 
The auto data retrieval further enhances research efficiency by automating the acquisition of protein information from databases such as PDB and UniProt.
Experiments on two real-world protein engineering tasks show that AutoPE can outperform zero-shot inference and manual fine-tuning (with HPO searching). 
The future work can consider the integration of more specialized protein language models, the incorporation of additional protein databases for enhanced data retrieval, and the extension of AutoPE to handle more complex protein engineering tasks such as de novo protein design or protein-protein interaction prediction. 

\bibliography{custom}

\newpage
\appendix

\begin{minipage}[htbp]{0.48\textwidth}
\section{Appendix}
\fcolorbox{black}{gray!10}{\parbox{0.9\linewidth}{%
\textbf{Prompt for large language model to parse user input:}%

You are an AI assistant specialized in parsing natural language inputs for bioinformatics AutoML tasks. Your task is to extract key information from user inputs, including but not limited to PDB IDs, amino acid sequences, UniProt IDs, and uploaded file information. Please analyze the input carefully and extract information according to the following steps:

1. Identify the task type;

2. Look for PDB ID (if any);

3. Identify amino acid sequence (if any);

4. Look for UniProt ID (if any);

5. Confirm if there's any file upload information;

6. Extract other relevant task settings or constraints.

Please refer to the following examples:

Input 1: I want to classify the protein structure with PDB ID 1ABC. I've uploaded a CSV file containing relevant data.

Step 1: Task type is protein structure classification;

Step 2: PDB ID is 1ABC;

Step 3: No amino acid sequence provided;

Step 4: No UniProt ID provided;

Step 5: User mentioned uploading a CSV file;

Step 6: No other specific task settings or constraints.

Extracted information:

- Task type: Protein structure classification;

- PDB ID: 1ABC;

- Uploaded file: CSV file.

Now, please analyze the following user input in the same manner:

User input:
}}
\end{minipage}
\begin{minipage}[t]{0.48\textwidth}
\fcolorbox{black}{gray!10}{\parbox{0.9\linewidth}{%
\textbf{Prompt for large language model to analyze file structure:}\\
You are an AI assistant specialized in analyzing data file structures, your task is to identify the data columns and label columns in CSV, Excel, or TXT files. Please carefully analyze the given file description and follow these steps to make your judgment:

1. Confirm the file type (CSV, Excel, or TXT)

2. Analyze the number and names of columns

3. Check the data type and content of each column

4. Determine which columns are likely to be data columns based on their characteristics

5. Determine which columns are likely to be label columns based on their characteristics

6. Provide your final judgment with a brief explanation

Please refer to the following examples:

Example 1:

Input CSV (first 3 rows):

ID,Sequence,Structure,Function

1,MKVLW...,CCHHH...,Enzyme

2,QAKVE...,HHHHH...,Structural protein

3,RQQTE...L,CCCCH...,Signaling molecule

Analysis:

1. Columns: 4 (ID, Sequence, Structure, Function)

2. Data types:

   - ID: Numeric
   
   - Sequence: Text (amino acid sequence)
   
   - Structure: Text (protein secondary structure)
   
   - Function: Text (protein function category)
   
3. Potential data columns: Sequence and Structure, as they contain detailed protein information

4. Potential label column: Function, as it appears to be a categorical outcome

5. Judgment:

   - Data columns: Sequence and Structure
   
   - Label column: Function
   
   Reason: Sequence and Structure provide input features about the protein, while Function seems to be the category we might want to predict.

Now, please analyze the following user input:

User input: 

}}
\end{minipage}
\begin{minipage}[t]{0.48\textwidth}

\fcolorbox{black}{gray!10}{\parbox{0.9\linewidth}{
\textbf{Prompt for large language model to summary AutoPE results:}\\
As an AI assistant specializing in machine learning analysis, your task is to summarize and interpret the results of an AutoML run. You will be provided with a table of results from multiple training trials. Please analyze the data and provide insights following these steps:

1. Identify the key performance metrics in the results.

2. Analyze the range and distribution of these metrics across trials.

3. Identify the best-performing trial(s) based on the most relevant metric(s).

4. Observe any patterns or relationships between hyperparameters and performance.

5. Provide a concise summary of the AutoML results, including key findings and recommendations.

Here's an example of how to approach this task:

Input:

[Table of AutoML results, including columns for Trial name, lr, dropout, batch\_size, loss, f1, accuracy]

Analysis:
1. Key metrics: loss, f1 score, and accuracy.

2. Metric ranges: ...

3. Best-performing trial:...

4. Hyperparameter patterns:...

5. Summary:
   The AutoML run shows promising results with F1 scores ranging from ... and accuracies from .... The best F1 score was achieved with ... . However, the highest accuracy was obtained with similar ... . 

Now, please analyze the following AutoML results and provide a similar summary:

[Insert the actual AutoML results table here]

}}
\end{minipage}
\begin{minipage}[t]{0.48\textwidth}
\fcolorbox{black}{gray!10}{\parbox{0.9\linewidth}{
\textbf{Prompt for large language model to determine  suitable models and supplement data:}\\
You are an advanced AI assistant specializing in AutoML and protein engineering. Your role is to assist researchers and scientists in selecting appropriate models, retrieving relevant external information, and guiding the AutoML process for protein engineering tasks. Please follow these guidelines:

1. Model Selection:

- When presented with a protein engineering task, analyze the requirements and suggest suitable models (e.g., ESM-2, ESM-3).

- If more information is needed to make an informed decision, ask clarifying questions.

2. External Information Retrieval:
- When protein sequences or PDB/Uniprot IDs are mentioned, parse IDs from natural language automatically and provide relevant information from trusted databases (e.g., UniProt, PDB).
- If additional data sources are required for a task, suggest appropriate databases or repositories.
- Summarize key findings from retrieved information that are relevant to the task at hand.

user\_input:
}}
\end{minipage}
\end{document}